\documentclass[a4paper,11pt]{article}
\usepackage{pos}

\newcommand{\bea}{\begin{eqnarray}}
\newcommand{\eea}{\end{eqnarray}}
\newcommand{\be}{\begin{equation}}
\newcommand{\ee}{\end{equation}}

\title{Axions, Black Holes and the Detection of Gravitons: from Astrophysics to Cosmology}
\ShortTitle{Axions, Black Holes and the detection of Gravitons}

\author*[a,b,c]{Nick E. Mavromatos}

\affiliation[a]{Department of Theoretical Physics and IFIC, University of Valencia and CSIC, E-46100, Valencia, Spain}

\affiliation[b] {Currently on leave from:
Physics Division, School of Applied Mathematics and Physical Sciences,
 National Technical University of Athens, Zografou Campus, 9 Iroon Polytechneiou Str., Athens 15780, Greece}
 
\affiliation[c]{Theoretical Particle Physics and Cosmology Group, Department of Physics, King's College London, Strand, London WC2R 2LS, UK}

\author[d]{Panagiotis Dorlis}

\affiliation[d] {
Physics Division, School of Applied Mathematics and Physical Sciences,
 National Technical University of Athens, Zografou Campus, 9 Iroon Polytechneiou Str., Athens 15780, Greece}

\author[b]{Sarben Sarkar}

\author[d]{Sotirios-Neilos Vlachos}

\emailAdd{mavroman@mail.ntua.gr}

\abstract{We review a novel scenario for the emergence of spin-polarisation entangled squeezed graviton states from superradiant axionic clouds in the neighborhood of 
astrophysical rotating black holes (BHs). The entangled squeezed graviton states are produced by both, conventional General-Relativity (GR) type axion-gravity interactions, and gravitational Chern-Simons (gCS)  anomalous terms coupled to axions, which are non-trivial in the presence of rotating BHs. The two kinds of terms have different-symmetry contributions to the entangled squeezed states. The squeezing parameter is estimated in a weak-quantum-gravity framework. Some phenomenology with respect to current and future interferometric detection devices is discussed. Importantly, current data from LIGO/Virgo Experiments can impose upper-bound constraints on the value of the squeezing parameter and, thus, on the lifetime of the axionic clouds. In addition to the above rather direct-detection possibility of squeezed gravitons, there is also the possibility of indirect detection of quantum gravitons in Cosmology, given that chiral quantum gravitational-wave (GW) perturbations in the primordial Universe may imply condensation of gCS terms. This, in turn, leads to inflation of running vacuum type, with in principle observable patterns in the profile of the GW produced during the post-inflationary early radiation  era, as well as the potential of alleviating cosmic tensions in the current era. 
\\

Preprint No.: {\bf KCL-PH-TH/2025-40}}

\FullConference{Proceedings of the Corfu Summer Institute 2025 "School and Workshops on Elementary Particle Physics and Gravity" (CORFU2025)\\
27 April - 28 September, 2025\\
Corfu, Greece\\}

\begin{document}

\maketitle

\section{Introduction}

Although General Relativity (GR), the classical field theory of the gravitational interaction, proposed by Einstein~\cite{Einstein:1915by}, works remarkably well at present, and has passed many important tests, including the observation of gravitational waves (GW) in 2015 by the LIGO-Virgo interferometers~\cite{LIGOScientific:2016aoc}, any experimental evidence about the quantum nature of gravity still eludes us. The question  whether gravity is a quantum theory, like the rest of the fundamental interactions in nature, has not yet been answered, despite considerable effort made, which lead to a diverse landscape of proposed theoretical approaches, ranging from string theory~\cite{str}, to loop quantum gravity~\cite{loop,loop2}, and asymptotic safety~\cite{as,as2}. There are even claims that gravity is not a fundamental but entropic force~\cite{Verlinde,Verlinde2}, emerging from the quantum entanglement of bits of information carried by the positions of material degrees of freedom.

One unambiguous path towards a verification of the quantum nature of the gravitational interaction (that is, of Quantum Gravity (QG)) would be the detection of gravitons, the alleged quantum-field theoretic carriers of the gravitational interaction. However, detecting single gravitons is a formidable unachieved task~\cite{Dyson:2013hbl}. In principle, it has been suggested that quantum fluctuations of the gravitational field could arise as {\it quantum} noise in interferometric searches~\cite{Amelino-Camelia:1998mjq,Amelino-Camelia:1999vks,Parikh:2020kfh}, whose characteristics depend on the quantum state of gravitons.  However, although interesting proposals have appeared recently in this direction~\cite{singlegrav}, it was pointed out~\cite{Carney:2023nzz} that such tests cannot distinguish between classical and quantum gravitational effects, given that similar effects can be produced by classical GW.

Such difficulties may be avoided if we look for novel \emph{collective} gravitational phenomena in which there might be a significant enhancement of  QG effects, making them accessible to experimental observation by current or future facilities. One such direction has been proposed in \cite{Dorlis:2025zzz,Dorlis:2025amf}, which we shall review here. The suggestion is to look for squeezed multi-mode quantum entangled states of gravitons produced 
by massive axion-like particles (ALPs) that may form 
appropriate condensates (``axionic clouds'') in the exterior of  rotating (Kerr-type) astrophysical BHs, characterised by an exponentially growing superradiant instability in the interaction of the ALPs with the BH~\cite{Detweiller,BritoCardoso}.
As argued in \cite{Dorlis:2025zzz,Dorlis:2025amf}, a consequence of the macroscopically large number of ALPs in the clouds, is a significant enhancement of the (collective) quantum effects present associated with quantum-entangled squeezed gravitons. These might lead to detectable signals in current or future interferometers~\cite{Berti:2005ys,Brito_2017_scales}. The massive axions will eventually decay to radiation, so the clouds will be characterised by a finite lifetime. Hence, the non-observation of squeezed gravitons by the currently-operating LIGO-Virgo-KARGA interferometer~\cite{LIGOScientific:2016aoc,McCuller:2021mbn} imposes a stringent upper bound on the GW-squeezing parameter 
$r < 41$~\cite{Hertzberg:2021rbl}, where  $\rm sinh^2 r \equiv \langle {\mathcal N}_{\rm sqgrav} \rangle$, with $\langle {\mathcal N}_{\rm sqgrav} \rangle$ denoting the average number of squeezed gravitons with respect to an appropriately excited quantum vacuum. As shown in \cite{Dorlis:2025zzz,Dorlis:2025amf}, and reviewed below, this in turn imposes an upper bound on the lifetime of the axion cloud, if the superradiance hypothesis is realised in Nature. 

Another framework where quantum gravitons can play an important r\^ole is inflation in (3+1)-dimensional modified gravity theories with gravitational anomalies~\cite{Basilakos:2019acj,Mavromatos:2020kzj,Mavromatos:2021urx,Dorlis:2024yqw,Dorlis:2024uei}, 
the so-called Chern-Simons (CS) gravity~\cite{Jackiw:2003pm,Alexander:2009tp}.
In such models, the gravitational CS (gCS) anomalous terms (which are total spacetime derivatives) couple to ALP fields. Unlike the interaction terms involving ALPs coupled to gauge CS anomaly terms, which are topological, the gCS-ALP interactions yield non-trivial contributions to the stress-tensor, and so to Einstein equations. In the cosmological scenario of \cite{Basilakos:2019acj,Mavromatos:2020kzj,Mavromatos:2021urx,Dorlis:2024yqw,Dorlis:2024uei}, at early epochs of the Universe, dominated by {\it chiral} (left-right-polarization asymmetric) GW can condense, and lead to gCS condensates. The latter imply a running-vacuum-model (RVM) type~\cite{SolaPeracaula:2022hpd,Lima:2013dmf} of inflation, which is probed by the non-linear gravitational terms in the effective vacuum energy density, which, in a cosmological setting, are proportional to the fourth power of the Hubble parameter $H^4$, or $H^4\, \ln H$~\cite{basilakos,Dorlis:2024yqw,Dorlis:2024uei}. The formation of a gCS condensate leads to an approximately linear ALP potential, as a result of the specific form of the gCS-ALP interaction term in CS gravity~\cite{Jackiw:2003pm,Alexander:2009tp}; this is reminiscent of the axion monodromy inflation, encountered in string cosmologies~\cite{silver}. However, in our case, unlike the case of \cite{silver}, inflation is mainly due to the $H^4$-terms in the condensate. Moreover, as demonstrated in \cite{Dorlis:2024uei}, the gCS condensates contain imaginary parts, which lead to a metastable inflation, with a finite lifetime. Agreement with the phenomenologically established order of 50 e-foldings~\cite{Planck}, then, can constrain the string scale~\cite{Dorlis:2024yqw}, in cases where the model is embedded in a string-inspired RVM framework~\cite{Basilakos:2019acj,Mavromatos:2020kzj,Mavromatos:2021urx,Dorlis:2024yqw,Dorlis:2024uei}. 

The important thing to note in the above inflationary scenario is that the chiral GW tensor perturbations are {\it quantum}. The gCS condensate has been estimated in \cite{Dorlis:2024yqw,Dorlis:2024uei} using a weak-graviton approximation, and as such, its value is approximate, and only within an effective field theory approach to quantum gravity~\cite{Donoghue:1994dn,Donoghue:2022eay}. The fact that the resulting inflationary ground state is of RVM type, implies prolonged reheating eras at the inflationary exit~\cite{Lima:2013dmf}. The latter may lead to an enhanced production of primordial BHs and the consequent presence of an early matter dominated (EMD) epoch~\cite{Carr:2018nkm}, which may lead to some distinctive effects on the profile of GW, detectable in future interferometers~\cite{Tzerefos:2024rgb}. Although the quantum effects of gravity in this case are not directly detectable, nonetheless the resulting RVM inflation and its prolonged reheating, including an early Dark Matter phase (eDM)~\cite{Carr:2018nkm}, is a direct consequence of them, due to its connection with the gCS condensate. 
Moreover, the effect of QG in such models is also to induce terms in the effective gravitational action logarithmic in spacetime curvature, which, in turn, leads to alleviation of current-era cosmological tensions~\cite{Gomez-Valent:2023hov}. In the above sense, the cosmological scenario of \cite{Basilakos:2019acj,Mavromatos:2020kzj}
provides another non-trivial example in which axions and quantum gravitons are linked.

The structure of the talk is the following: in the next section \ref{sec:2}, we review the proposed basic mechanism of the production of squeezed graviton states.
In section \ref{sec:3}, we estimate the squeezing parameter, by the total number of squeezed gravitons and discuss the imposed upper limit of the axionic cloud by the current non-observation of squeezed graviton states. 
In section \ref{sec:qggw}
 we discuss principles of detection of quantum features in gravitational waves of relevance to observational tests of our discussion here.
Section \ref{sec:4} describes briefly the basic features of the cosmological RVM-type inflationary model of \cite{Basilakos:2019acj,Mavromatos:2020kzj}, and its distinctive features from GR, as far as the profiles of GW during radiation era, or deviations from $\Lambda$CDM Cosmology in modern eras, and potential alleviations of the cosmological tensions, are concerned.
Finally section \ref{sec:5} contains our conclusions and outlook. In Appendix \ref{sec:app} we review, for completeness, consequences of the non-observation of single-mode squeezed graviton states by current interferometers. Specifically, we explain how this 
leads to the imposition of an allowed upper bound of the squeezing parameter, which is used in section \ref{sec:3} to exclude phenomenologically  very long lifetimes of ALP clouds.

\section{Superradiant axionic clouds and quantum-entangled squeezed graviton states}\label{sec:2}

Superradiance~\cite{zel1,zel2} is a known phenomenon in electromagnetism, during which the amplitude of the transmitted electromagnetic wave in interaction with a rotating object is larger than that of the incident wave, thus leading to amplification, provided the radiation is prepared in an appropriate angular momentum state.
Specifically, if $\omega_r$ is the frequency of radiation, $m$ the eigenvalue of the component of its angular momentum along the axis of rotation of the object, and $\Omega$ the frequency of the rotating object, then to have superradiance the following condition should occur:
\begin{align}\label{srcond}
    \omega_r < m\, \Omega\,.
\end{align}

Superradiance also characterises a massive (pseudo)scalar matter field in interaction with the background of a rotating BH~\cite{Detweiller,BritoCardoso}, which is the case of interest in this talk, and in \cite{Dorlis:2025zzz,Dorlis:2025amf}, for which the matter particle is an ALP. In such a case there is an exponentially growing instability in the interaction between the ALP and the BH, which leads to a production of a large number of squeezed quantum-entangled gravitons as a result of either annihilations or decays of the axion fields. The condition \eqref{srcond} is also valid intact in that case, with $\omega_r \to \omega $ corresponding to the real part of the frequency of the ALP (pseudoscalar) field mode, and $\Omega \to \Omega_{H}$ representing the angular velocity of the rotating BH outer event horizon. 

\begin{figure}[ht]
    \centering
    \includegraphics[width=0.6\linewidth]{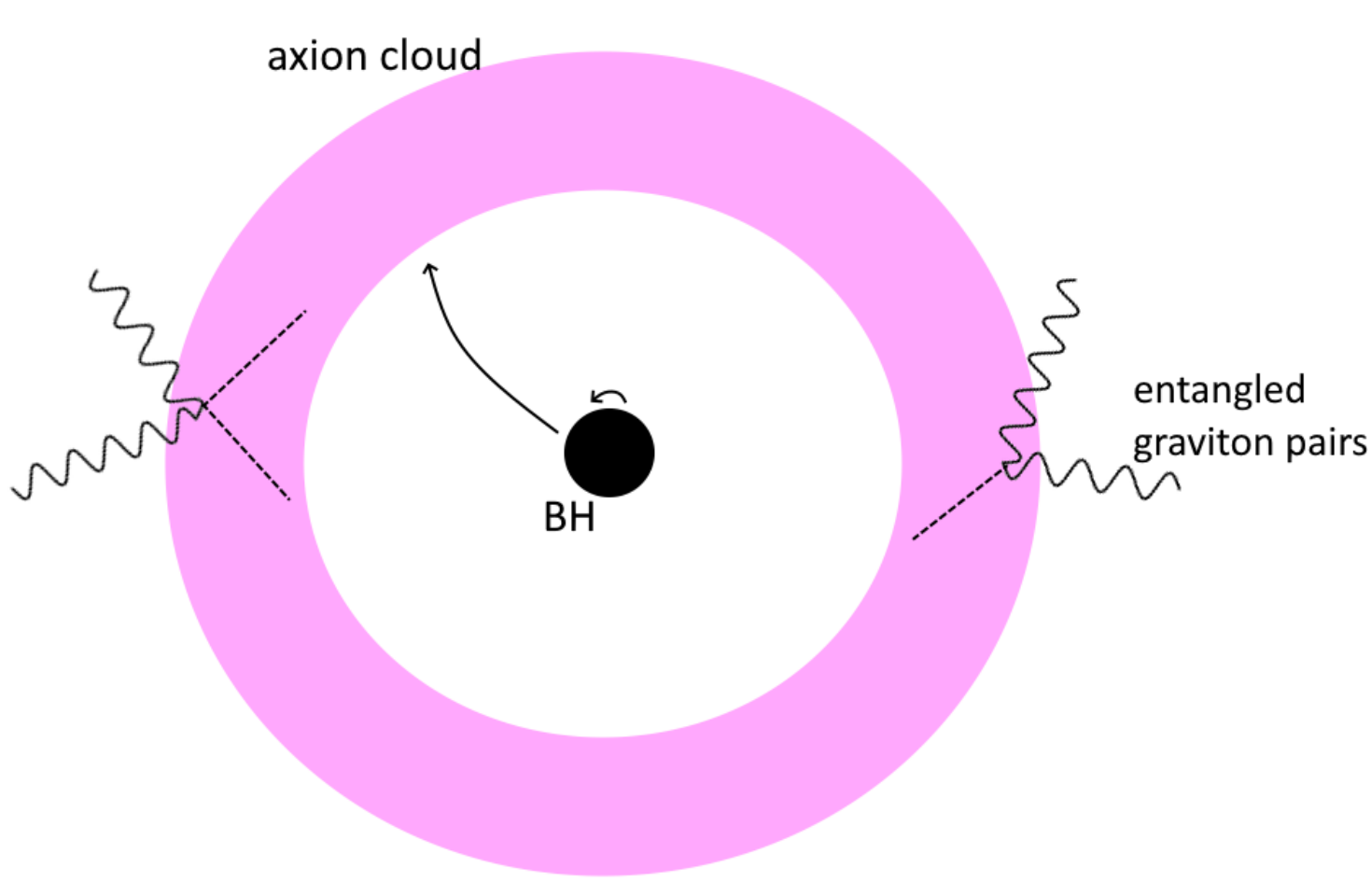}
    \caption{Superradiant axionic cloud around a rotating black hole (dark blob). The non-linear interactions of axions (dashed lines) with gravitons (wavy lines) producing quantum-entangled squeezed graviton pairs are indicated explicitly in a weak-gravity effective field theory framework, we restrict our attention for our purposes here. Picture taken from \cite{Dorlis:2025zzz}.}
    \label{cloud}
\end{figure}

\begin{figure}
    \centering
\includegraphics[width=0.7\linewidth]{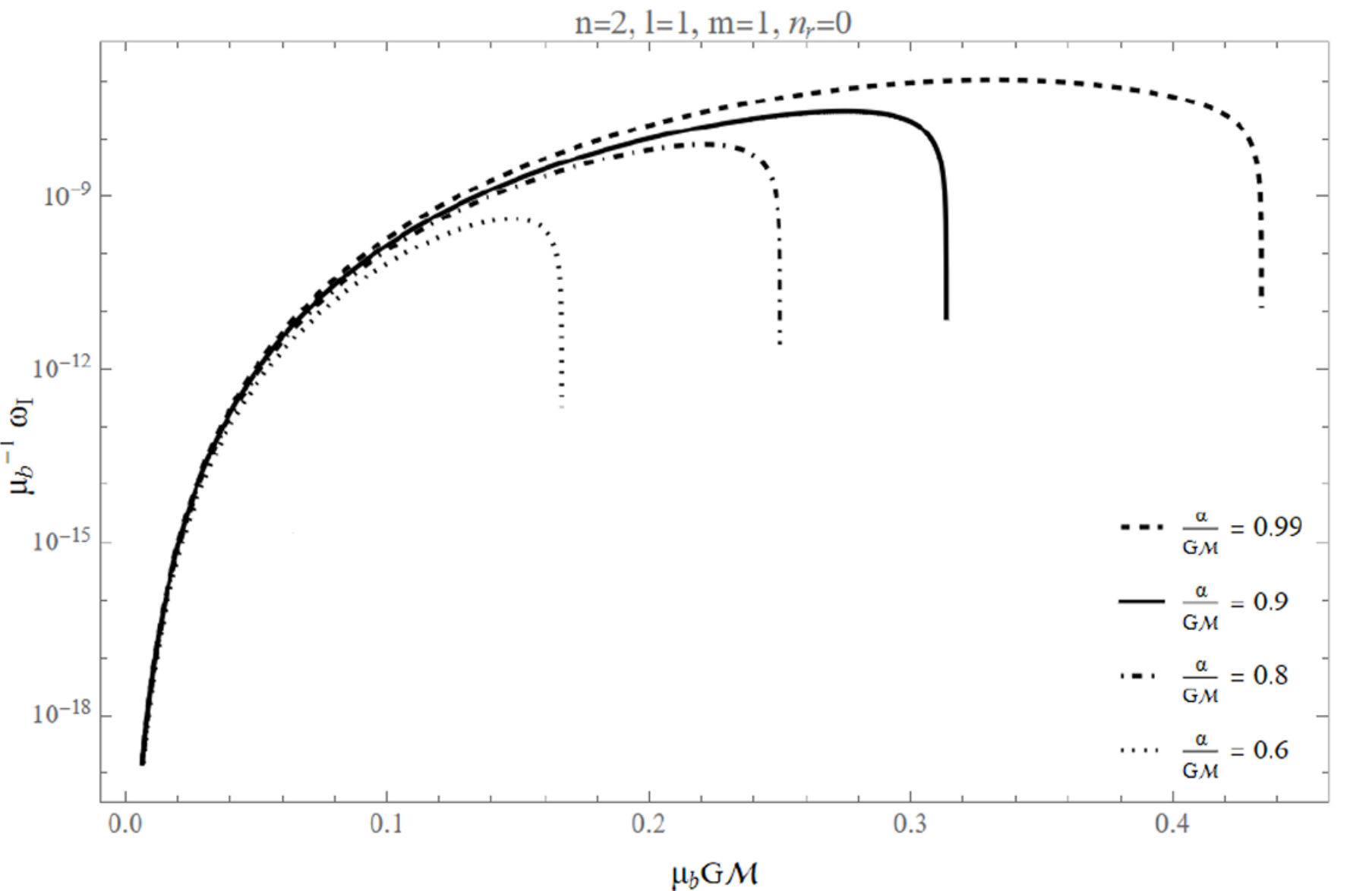}
    \caption{The growth rate stemming from the superradiant instability of the dominant ''$2p$-state" in the axion cloud of the rotating black hole of figure~\ref{cloud}, for different values of the BH spin ratio $\alpha/(G \mathcal{M})$. Picture taken from \cite{Dorlis:2025amf}.}
    \label{growth_rate}
\end{figure}

In a background of a rotating (Kerr-type~\cite{Kerr:1963ud}) BH of mass $\mathcal{M}$, angular momentum  $\mathcal{J}_H$ and dimensionless spin parameter $ \alpha = \mathcal{J}_H/\mathcal{M}$, the Klein-Gordon equation (KG)  for a massive (pseudo)scalar field, of mass $\mu_b$, admits quasibound state solutions~\cite{Detweiller}, labeled by integers $\left(n,l,m\right)$. 
The ALP-BH system develops an exponentially-growing instability when the pseudoscalar field frequency acquires a {\it positive} imaginary part $\omega_I >0$~\cite{Detweiller,BritoCardoso}. This drives the superradiant instability with a growth $\sim e^{\omega_I t}$, provided the real part of the frequency, $
\omega$, satisfies  the superradiant condition \eqref{srcond}~\cite{Bekenstein:1973mi}. We have $\omega \approx \mu_b\left(1 - \frac{a_{\mu}^2}{2n^2}\right)$, where $a_\mu \equiv G \mathcal{M}\mu_b\,$ (with $\mu_b$ the ALP mass and $\mathcal M$ the BH mass) is  the dimensionless coupling of the gravitational atom. In \cite{Dorlis:2025zzz,Dorlis:2025amf} we worked in the non-relativistic regime~\cite{Detweiller}, 
for which the Compton wavelength of the ALP $\lambda = \mu_b^{-1} \gg G\mathcal M$, implying $a_\mu \ll 1$. In this approximation the classical ALP field is given by:
\begin{equation}\label{classALP}
    b(t,r,\theta,\phi)=\sum_{nlm}e^{-i \omega_{n\ell m} t} \sqrt{\frac{N_{nlm}}{2\mu_b}}\Psi_{nlm} + \text{c.c} \ ,
\end{equation}
where c.c. denoting complex conjugation, and $N_{nlm}$ is the number of axions in the respective state, with $\Psi_{nlm}(\vec{x})$ obeying the  normalization $\int d^3x \vert \Psi_{nlm}\vert^2
=1$. As long as the superradiance condition \eqref{srcond} is satisfied for the real part of the ALP frequency $\omega$, the axionic-cloud will grow at a rate faster than the evolution timescale of the BH. The classical field \eqref{classALP} characterises the axionic cloud, which is a kind of condensate of finite life time, due to the eventual decay of the massive ALPs. Such condensates are characterised by a large number $\vert N_{n\ell m} \vert \gg 1$, which is eventually responsible for the significant enhancement of the number of the produced squeezed graviton states as a result of the fusion or decay processes of ALPs ({\it cf.} figure \ref{cloud}). For our purposes we note that the most dominant mode corresponds to the ``$2p$-axion state" $(n=2, \ l=m=1)$, on which we concentrate our attention here. The superradiant instability growth of this state is depicted in figure \ref{growth_rate}.

The important point to realize is that, in the presence of ALPs $b(x)$ fields interacting with a rotating BH background, the gCS anomaly term: 
\begin{align}\label{Rcs}
    S_{CS}  = -\frac{A}{2}\,\int d^{4}x\sqrt{-g}\ \ b\ \, R_{\mu\nu\rho\sigma}\widetilde{R}^{\nu\mu\rho\sigma} \  , 
\end{align}
where $A$ denotes the coupling constant, and $\widetilde{R}_{\mu\nu\rho\sigma}=\frac{1}{2}R_{\mu\nu\alpha\beta}\ \epsilon^{\alpha\beta}\!_{\rho\sigma}\ $ is the dual of the Riemann tensor, with  $\epsilon_{\mu\nu\rho\sigma}$ the covariant Levi-Civita tensor, is non trivial, thereby implying an effective coupling described by a CS gravity action~\cite{Jackiw:2003pm}:\footnote{Our conventions and definitions used throughout this work are: signature of metric $(-, +,+,+ )$, Riemann Curvature tensor:
$R^\lambda_{\,\,\,\,\mu \nu \sigma} = \partial_\nu \, \Gamma^\lambda_{\,\,\mu\sigma} + \Gamma^\rho_{\,\, \mu\sigma} \, \Gamma^\lambda_{\,\, \rho\nu} - (\nu \leftrightarrow \sigma)$, Ricci tensor $R_{\mu\nu} = R^\lambda_{\,\,\,\,\mu \lambda \nu}$, and Ricci scalar $R_{\mu\nu}g^{\mu\nu}$. We also work in units $\hbar=c=1$.}
\begin{align}\label{CSgravaction}
S = \frac{1}{2\kappa^2} \int d^4x \, \sqrt{-g}\, R + S_{CS} + S_m\,,    
\end{align}
where  $\kappa = M_{\rm Pl}^{-1}$ is the (3+1)-dimensional gravitational constant, with 
$M_{\rm Pl} =  2.435 \times 10^{18}$~GeV the (reduced) Planck mass scale and $S_m$ denotes the matter action, including the dynamics of the massive ALP field $b(x)$, 
\begin{align}\label{matter}
S_m = \frac{1}{2} \partial_\mu\, b\partial^\mu b - \frac{1}{2} \mu_b^2 \, b^2 + \dots \,.     
\end{align}
The $\dots$ denotes possible interactions of the ALP field, as well as other matter (or radiation) fields, which are not of direct relevance to us here. If the model is embedded in string theory, the coupling $A$ is given by~\cite{Duncan:1992vz}:
\begin{align}
\label{Adef}
A= 
  \sqrt{\frac{2}{3}}\frac{M_{\rm Pl}}{48\, M_s^2} \,,
\end{align}
where $M_s$ is the string mass scale~\cite{str1}.
In \cite{Dorlis:2025zzz,Dorlis:2025amf}, a weak graviton approach has been adopted, in which one expands the graviton field about the rotating (Kerr-type) BH background $g_{\mu\nu}^{(0)}$:
\begin{align}\label{metrpert}
g_{\mu\nu} = g_{\mu\nu}^{(0)} + \kappa h_{\mu\nu}\,, 
\end{align}
where  $\kappa \vert h_{\mu\nu} \vert \ll 1$, is the metric (GW type) perturbation in the transverse and traceless (TT) gauge. In our approach, the bulk of the produced squeezed gravitons is produced in regions of the cloud, which, in the non-relativistic approximation, lie far away from the outer horizon of the BH~\cite{BritoCardoso}; hence the spacetime metric can be well approximated by the Minkowski flat metric $g_{\mu\nu}^{(0)} \simeq \eta_{\mu\nu}$. Upon substituting  the expansion \eqref{metrpert} onto the action \eqref{CSgravaction}, and truncating the weak-graviton expansion to second order in tensor perturbations $h_{\mu\nu}$, we obtain the following axion-graviton interactions, in the TT-gauge:
\begin{align}\label{S12}
    S^{(1)}&= \frac{\kappa}{2}\int d^4x \ h_{ij}T^{ij}  \ ,  
    \nonumber \\ 
    S^{(2)}&=  - \frac{\kappa^2}{2} \int d^4x \ h_{im}h^{m}_{\ j} \ \partial^i b\,\partial^j b \ ,  \nonumber \\
     S^{(2)}_{CS} & = -A\kappa^2\int d^4x \ b(x) \epsilon_{ijk} \ \Bigg( \partial_l\partial^k  h^j_m  \partial^m\dot{h}^{li}  +\ddot{h}^{li}\partial^k\dot{h}^j_l  
    -\partial_m\partial^kh^j_l\partial^m\dot{h}^{li}  \Bigg)   
\end{align}
 where $T_{ij}$ is the matter stress-momentum tensor (Latin indices are 3-space ones, $i, j, \dots = 1,2,3)$.

Going to Fourier space, one can {\it quantize} the system of graviton perturbations $h_{ij}$ in an initially finite volume $V$:
\begin{equation}
    \hat{h}_{ij}(t,\vec{x})=M_{\rm Pl}\sum_{\vec{k},\lambda} f_k\left[   e^{(\lambda)}_{ij}(\vec{k}) \hat{\alpha}^\dagger_{\lambda,\vec{k}}\ e^{-ik\cdot x} +h.c  .  \right] \ ,
    \label{mode_expansion_gravitons_}
\end{equation}
where $\hat{\alpha}_{\lambda,\vec{k}},\hat{\alpha}^{\dagger}_{\lambda,\vec{k}}$ denotes the  dimensionless annihilation/creation operators obeying the usual commutation relations,
$ [\hat{\alpha}_{\lambda,\vec{k}}  ,  \hat{\alpha}^\dagger_{\lambda^\prime , \vec{k}^\prime}] = \delta_{\lambda\lambda^\prime} \delta_{\vec{k}\vec{k}^\prime}$, where $\lambda=L,R$ denotes left (L) and right (R) polarisation respectively,  $f_k$ is the (dimensionless) single graviton  strain, $f_k\equiv \kappa(2V\Omega_k)^{-\frac{1}{2}}$, while $\Omega_{k}\equiv k$ is the frequency of GW with momentum $\vec k$. The polarisation tensors $e_{(\lambda^\prime)}^{ij}$ are subject to normalization
$e^{*(\lambda)}_{ij} e_{(\lambda^\prime)}^{ij}=2\delta_{\lambda\lambda^\prime}$. In regions of the axion cloud far away from the BH horizon ({\it cf.} figure~\ref{cloud}) the quantum vacuum on which the  graviton fields ${\hat h}_{ij}$ act  is, to a good approximation,  the Minkowski spacetime vacuum; we  restrict our attention to this case  from now on.

From the first term interaction term $S^{(1)}$ in \eqref{S12}, we obtain coherent gravitons, upon quantisation of the tensor perturbations. These are not of interest to us here. The terms we are really interested in, which yield squeezed entangled graviton states  are the purely GR term $S^{(2)}$, which describes thew fusion of two ALP fields into a pair of gravitons, and the gCS term $S^{(3)}$, which pertains to a decay of an ALP into two gravitons. The squeezed graviton quantum operator is defined in analogy with the quantum optics case~\cite{Scully_Zubairy_1997,Agarwal_2012}:
\begin{equation}
\label{Evolution_Operator}
    \hat{S}=\exp\left[ \frac{1}{2}\sum_{I,J}\mathcal{G}_{IJ}\ \hat{\alpha}^\dagger_I\hat{\alpha}^\dagger_J -h.c.    \right]
\end{equation}
where the index $I=(\lambda,\vec{k})$ denotes the graviton states, and 
\begin{equation}
\label{G_IJ}
    \mathcal{G}_{IJ}=-2\ i\ \mathcal{F}_{IJ} \  T \ \text{sinc}\left[\left(\Omega_k + \Omega_{k^\prime} - E \right) \frac{T}{2}\right] 
\end{equation}
is the analogue of the multi-mode squeezing parameter in our BH-ALP system; the quantity $T$ is the lifetime of the axionic condensate (cloud) around the BH, which plays the r\^ole of the classical (coherent) source driving the process. The interaction coefficients $\mathcal F_{IJ}$ appearing in \eqref{G_IJ}, are defined through the generic structure of the interaction Hamiltonian for the GW (tensor) perturbations~\cite{Dorlis:2025zzz,Dorlis:2025zzz}:
\begin{equation}
\label{Hamiltonian_general_structure}
\begin{aligned}
    \hat{H}_{\text{int}} =\sum_{I,J}e^{i(\Omega_k+\Omega_{k^\prime}-E)t}  \mathcal{F}_{IJ} \hat{\alpha}^\dagger_I\hat{\alpha}^\dagger_J+  h.c  +  \dots 
    \end{aligned}
\end{equation}
 while $E$ is the axion energy available for the production of gravitons in the appropriate processes indicated in  \eqref{S12}. In the adopted rotating-wave approximation (RWA) \cite{WallsMilburn2008,PhysRevA.31.2409,Wu:1986zz,SFWM1,SFWM2,SFWM_3}, terms with mixed graviton creation and annihilation operators (denoted by $\dots$) are ignored, given that they are subdominant for long enough interaction times we are considering here, because they oscillate rapidly. In our case, for the GR-induced interaction $hhbb$ ($S^{(2)}$ in \eqref{S12}), we have: 
\begin{equation}
\mathcal{F}^{(GR)}_{(\vec{k},\lambda)(\vec{k}^\prime,\lambda^\prime)} = \frac{  f_k f_{k^\prime} N_{2p}}{4\mu_b} \mathcal{I}^{(GR)}_{(\lambda,\vec{k})(\lambda^\prime,\vec{k}^\prime)} \, ,
\label{F_GR}
\end{equation}
where 
\begin{equation}
\label{I_GR_Correlation}
     \mathcal{I}^{(GR)}_{(\lambda,\vec{k})(\lambda^\prime,\vec{k}^\prime)}=e^{(\lambda)}_{im}(\vec{k}) I_{ij}(\vec{k},\vec{k}^\prime \ )e^{(\lambda^\prime)}_{mj}(\vec{k}^\prime)\,,
\end{equation}
and $I_{ij}$ has the tensorial structure  ,
\begin{equation}\label{Iij}
    I_{ij}(\vec{k},\vec{k}^\prime)=\int d^{3} \vec{x} \ \partial_i \Psi_{2p}(\vec{x})\partial_j \Psi_{2p}(\vec{x})e^{-i(\vec{k}+\vec{k}^\prime)\cdot \vec{x}}\, ,
\end{equation}
with $E=2\mu_b$, since two axions are involved in the process.
The axion occupation number $N_{2p}$ acts as an enhancement parameter, as already mentioned. This process is the gravitational analogue of the spontaneous four-wave mixing (SFWM) in quantum optics~\cite{SFWM1,SFWM2,SFWM_3}, in which two ``pump’’ photons interact inside a nonlinear medium, resulting in the production of quantum-entangled photon pairs. The so-produced graviton pairs are also entangled, as they belong to a non-separable state~\cite{Dorlis:2025zzz,Dorlis:2025zzz}. 

On the other hand, in the gCS anomalous interaction $S^{(2)}_{CS}$ in \eqref{S12}, involving one ALP and two gravitons, 
the relevant kernels are given by:
\begin{equation}\label{IijFour}
   \mathcal{F}^{(CS)}_{(\lambda,\vec{k})(\lambda^\prime,\vec{k}^\prime)}=   iA\sqrt{\frac{N_{2p}}{2\mu_b}}f_k f_{k^\prime} \ \Omega^{2}_k 
 \ \Omega^{2}_{k^\prime}\ \mathcal{I}^{(CS)}_{(\lambda,\vec{k})(\lambda^\prime,\vec{k}^\prime)} \ ,
\end{equation}
where 
\begin{equation}
\label{I_CS_Correlations}
\begin{aligned}
    &\mathcal{I}^{(CS)}_{(\lambda,\vec{k})(\lambda^\prime,\vec{k}^\prime)}=l_{\vec{k}^\prime}l_{\lambda^\prime}\widetilde{\Psi}_{2p}(-\vec{k}-\vec{k}^{\prime})\times\\
    &\times\left(   [e^{(3)}(\vec{k}^\prime)] _m  \ e^{(\lambda)}_{mj}(\vec{k})\  e^{(\lambda^\prime)}_{jl}(\vec{k}^\prime) [e^{(3)}(\vec{k})] _l 
   \  +\right.\\
    &\left. + \left(1-   \cos\Delta\theta \right)  \ e^{(\lambda)}_{mj}(\vec{k}) \   e^{(\lambda^\prime)}_{mj}(\vec{k}^\prime)      \     \right) \ ,
    \end{aligned}
\end{equation}
with $e^{(3)}( \vec{k} ) =\vec{k}/| \vec{k} |$, and $\widetilde{\Psi}_{2p}(\vec{k})$ denotes the Fourier transform of $\Psi_{2p}(\vec{x})$, $l_\lambda=\pm1$, for $\lambda=L,R$ and $l_{\vec{k}}=\pm 1$, for $\theta_k>\pi/2$ or $\theta_k<\pi/2$, respectively, 
where $\theta_k$ is the polar angle of $\vec{k}$ \cite{Alexander:2004wk}.  Here  $E=\mu_b$ and the proportionality factor is $\sqrt{N_{2p}}$, given that  only one axion is involved in the microscopic process. The quantum-optics analogue of this process is the so-called Spontaneous parametric down-conversion (SPDC)~\cite{PhysRevA.31.2409, Wu:1986zz}, in which one high-energy (``pump’’) photon splits into two lower-energy entangled photons \cite{SPDC_multimode}.

\section{Estimates of the squeezing parameter and   axionic-cloud-lifetime constraints}\label{sec:3}

The squeezing operator \eqref{Evolution_Operator} has the property~\cite{multimode}: 
\begin{align}
\label{SaS_multimode}
    \hat{S^\dagger} \hat{\alpha}_{I} \hat{S} =\sum_J (\mu_{IJ}\hat{\alpha}_J  +  \nu_{IJ}\hat{\alpha}^\dagger_{J})\ , 
\end{align}
where:
\begin{align}
\label{mu_IJ_expansion} 
\mu_{IJ}=\delta_{IJ} + \frac{1}{2!}\sum_M \mathcal{G}_{IM}\mathcal{G}^\star_{MJ} \ + \ \dots \,,
\end{align}
\begin{align} 
\label{nu_IJ_expansion}
\nu_{IJ} = \mathcal{G}_{IJ} + \frac{1}{3!}\sum_{M,L} \mathcal{G}_{IM}\mathcal{G}^\star_{ML} \mathcal{G}_{LJ} \ + \ \dots \,  \ .
\end{align}
Then, the average number of gravitons $N_{gr}$ in the squeezed vacuum state $|\psi\rangle=\hat{S}\vert 0 \rangle $ can be constrained by:
\begin{align}\label{Nmax}
\langle N_{gr} \rangle = \sum_{I, J} 
\Big|\nu_{IJ}\Big|^2\ \lesssim \sum_{I,J} \Big( \Big|{\mathcal G}_{IJ}\Big|^2 \ + \  \dots \Big)\,,
\end{align}
where we have used \eqref{nu_IJ_expansion}, the $\dots$ denote the contributions from the cubic and higher-order terms in \eqref{nu_IJ_expansion}, 
and, for the upper bound, we use the triangle inequality.
In the case of a single-mode squeezed vacuum~\cite{Guerreiro:2019vbq}, $\mathcal{G}_{IJ}\sim\delta_{IJ}$, and the expression \eqref{Nmax} reduces to the well known relation 
\begin{align}\label{singlemode}
\langle N_{gr}\rangle=\sinh^2 r\,. 
\end{align}
To have good observational prospects of squeezed gravitons, which would constitute one of the most decisive tests of QG, 
the first term of the infinite series \eqref{Nmax} (or \eqref{nu_IJ_expansion}) must be of order one or higher. In such a case, all terms in the series should be resumed, leading to an exponential enhancement of the squeezed gravitons, similar to the single mode case \eqref{singlemode}~\cite{Dorlis:2025zzz,Dorlis:2025amf}.  
\begin{figure}[ht]
    \centering
    \includegraphics[width=0.5\linewidth]{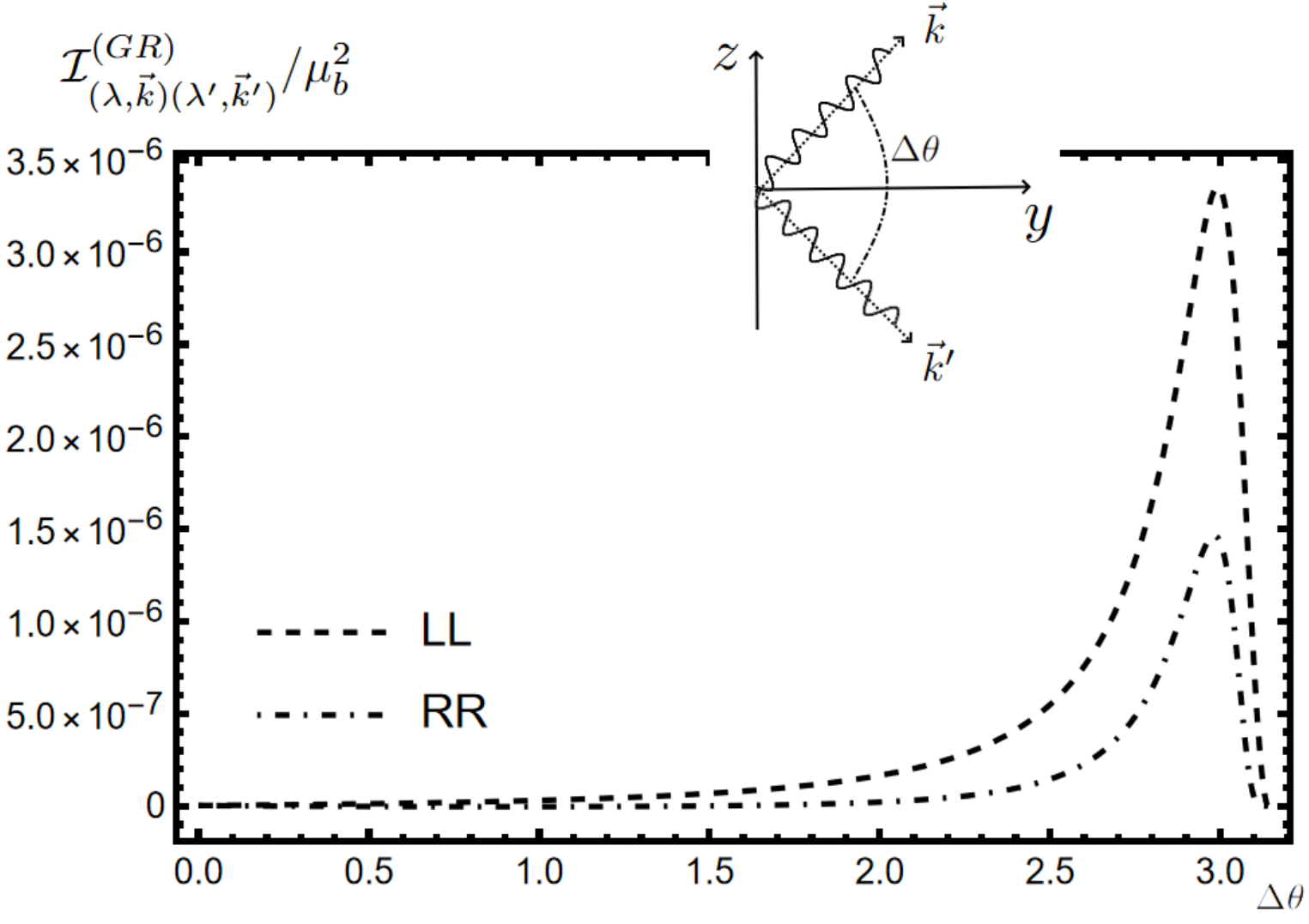}\hfill \includegraphics[width=0.5\linewidth]{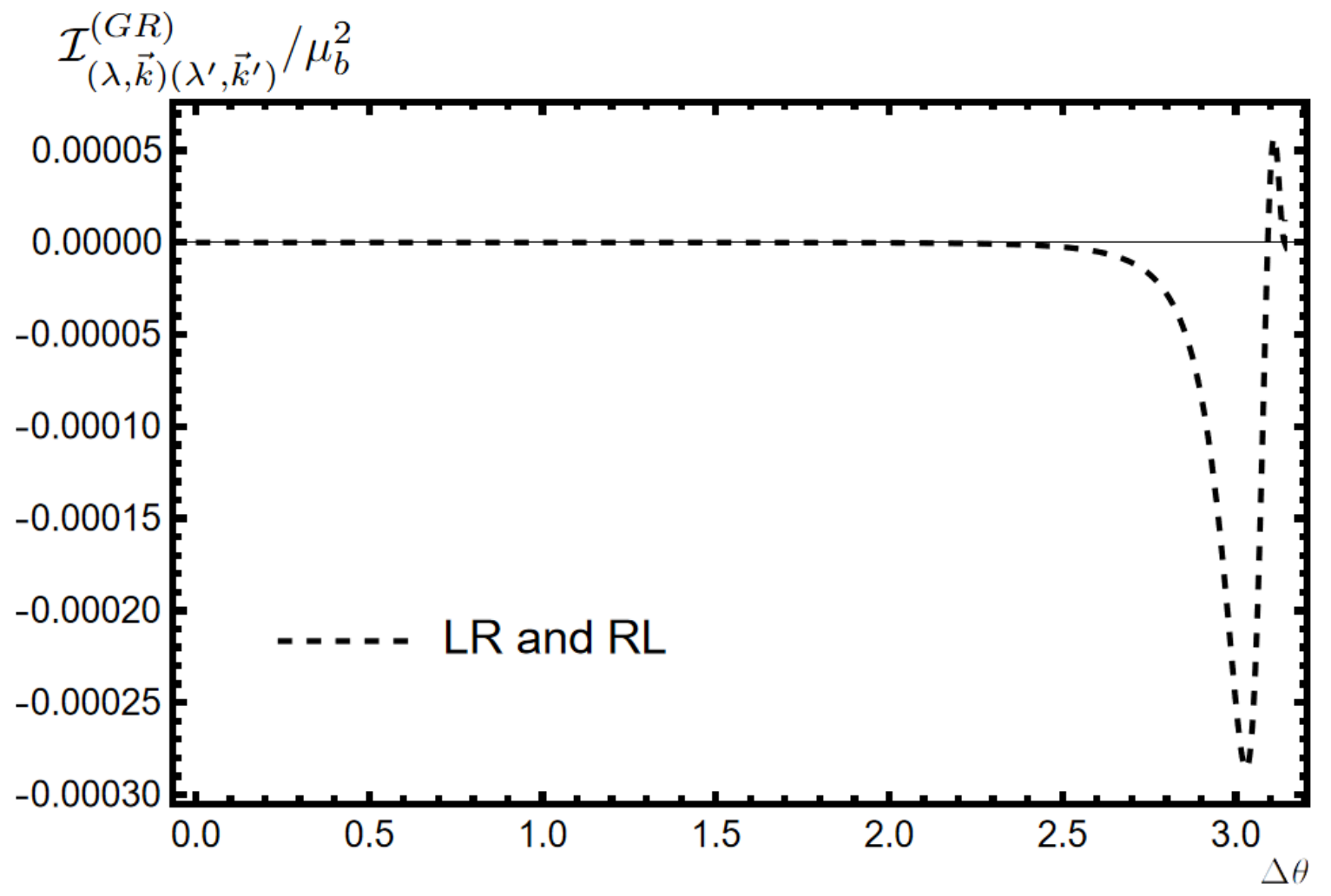}
    \caption{Angular and polarisation correlations for the GR interaction $S^{(2)}$ in \eqref{S12}. The $2p$-state results in an asymmetry between the LL and RR pairs. The above figures correspond to $a_\mu=0.1$. Picture taken from Ref.~\cite{Dorlis:2025zzz}.}
    \label{GR_correlations}
\end{figure}

For the GR interaction $S^{(2)}$ in \eqref{S12}, we consider the case of figure \ref{GR_correlations}.
From \eqref{G_IJ}, \eqref{F_GR}, \eqref{I_GR_Correlation}, \eqref{Iij} and \eqref{Nmax}, we find the following differential expression per solid angles~\cite{Dorlis:2025zzz,Dorlis:2025amf}:
\begin{equation}
\begin{aligned}
    \frac{d^2}{d\Omega d\Omega^\prime}\sum_{I,J} \Big|(\mathcal{G}^{(GR)}_{IJ})\Big|^2=&\frac{1}{128\pi^3}\left(\frac{\mu_b}{M_{\rm Pl}}\right)^{4}N^{2}_{2p}\,(T \mu_b )\times
\\
&\times\sum_{\lambda,\lambda^\prime}\int d\tilde{k}d\tilde{k}^\prime \ \tilde{k}\tilde{k}^\prime\delta\left(
    \tilde{k}+ \tilde{k}^\prime -2\right)\Big | \widetilde{\mathcal{I}}^{(GR)}_{(\lambda,\vec{k})(\lambda^\prime,\vec{k}^\prime)}\Big|^{2} \ ,
    \end{aligned}
    \label{N_gr_per_Solid_angles}
\end{equation}
where $\tilde{k}^{(\prime)} \equiv k^{(\prime)}/\mu_b$, $\mathcal{I} \equiv \widetilde{\mathcal{I}}/\mu_b^2$. 
In order to estimate the number of gravitons in the squeezed vacuum state, we consider $\vec{q}=-(\vec{k}+\vec{k}^\prime)$ lying on $\theta_{k+k^\prime}=\varphi_{k+k^\prime}=\pi/2$ ({\it c.f.} figure \ref{GR_correlations}). Taking into account that the main contribution
to the integral \eqref{N_gr_per_Solid_angles} comes from momentum vectors $\vec k$, $\vec k^\prime$  satisfying $k\approx k^\prime\approx\mu_b$, whilst the maximum contribution corresponds to their relative angle $\Delta \theta=0.964\pi$, then, after integrations over the solid angles, we obtain 
a factor at most of 
order $\left(4 \pi \right)^2$, implying the following upper limit for the GR-related squeezing quantity~\cite{Dorlis:2025zzz,Dorlis:2025amf}:
\begin{equation}
    \label{N_gr_RESULT}
\sum_{I,J}\vert\mathcal{G}^{(GR)}_{IJ}\vert^2 \lesssim 2.5\times10^{-15} \,T \mu_b\, ,
\end{equation}
where $T$ can be identified~\cite{Dorlis:2025zzz,Dorlis:2025amf} with the lifetime of the axionic cloud surrounding the BH ({\it cf.} figure~\ref{cloud}).
The reader should observe our main result, that the large suppression induced by the ratio $\big(\mu_b/M_{\rm Pl}\big)$ in \eqref{N_gr_per_Solid_angles} is compensated by the number of axions in the cloud ,  
$ N_{2p}\left( \mu_b/M_{\rm Pl}\right)^2\approx 10^{-3}a_\mu^2 =10^{-5}\ $, bringing QG effects of the EFT approach into observationally promising values.\par 

\begin{figure}[ht]
    \centering
    \includegraphics[width=0.7\linewidth]{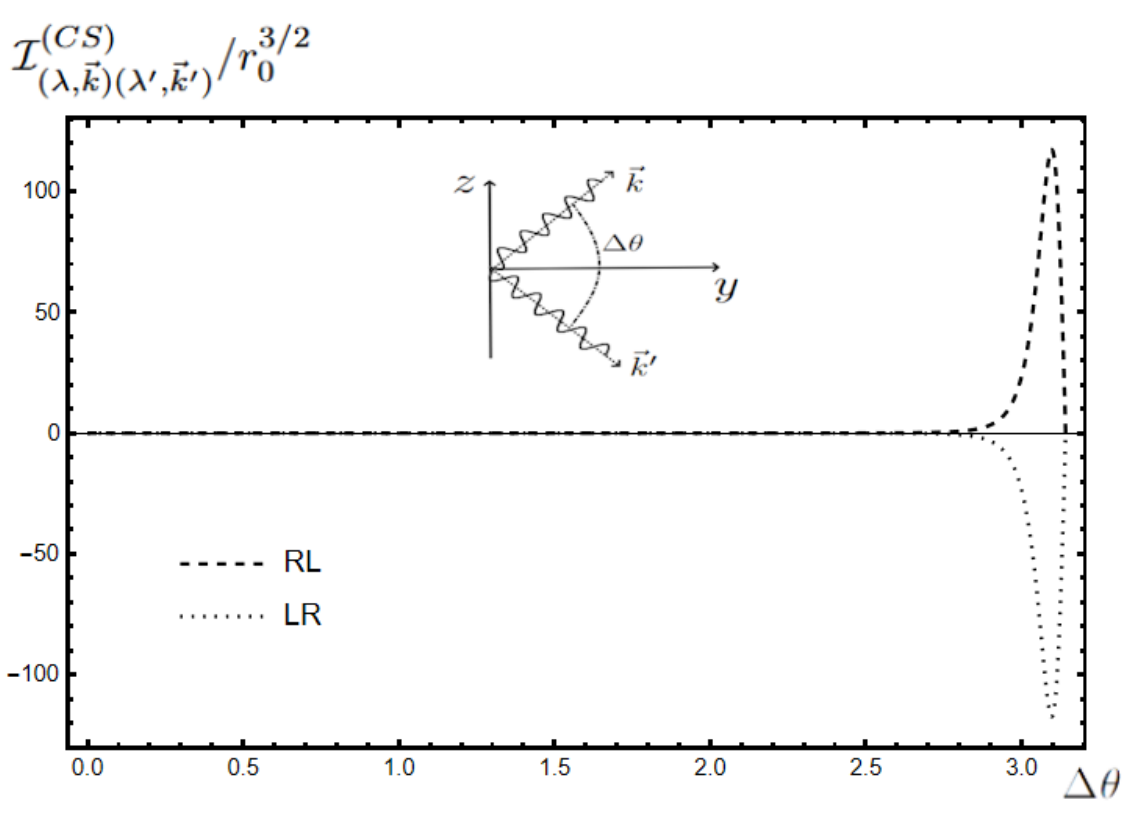}
    \caption{Angular and polarisation correlations for the anomalous gravitational CS interaction $S^{(2)}_{CS}$ in \eqref{S12}. Only pairs of opposite polarisations are produced, with Maximal entanglement occurring between the L and R polarisations. The plot corresponds to $a_\mu=0.1$. Picture taken from \cite{Dorlis:2025zzz}.}
    \label{CS_correlations}
\end{figure}

This is to be contrasted with the case of the anomaly-induced interaction $S^{(2)}_{CS}$ in \eqref{S12}. In an effective field theory, the CS-coupling to the axion is suppressed below the cut-off scale of the theory.  
The pertinent squeezing effect is proportional to the coupling strength of the axion to the (gravitational) CS anomaly.
For the string-inspired KR axion coupling \cite{Duncan:1992vz} $A\sim 10^{-2}M_{\rm Pl}/M^{2}_\text{s}$, with $M_\text{s}$ the string scale.   As in the GR case, we argue ({\it c.f.} figure~\ref{CS_correlations}):
\begin{equation}
\label{N_cs_RESULT}
    \sum_{I,J} \Big|(\mathcal{G}^{(CS)}_{IJ})\Big|^2 \lesssim 10^{-10}\left(\frac{\mu_b}{M_s}\right)^4 \, \mu_b T\ .
\end{equation}

It is evident from \eqref{N_gr_RESULT} and \eqref{N_cs_RESULT} that the lifetime  $T$  of the axionic cloud is  also a  parameter able to induce significant squeezing. In contrast to the short lived quasi-normal modes (QNMs) \cite{BH_Squeezer}, the longevity of axionic clouds seems to have the capability of producing appreciable squeezing effects, provided is sufficiently large. 

The lifetime of the axionic condensate has been estimated by various groups~\cite{BritoScales1,BritoScales2,Yoshino_2014_scales,porto_scales,BritoCardoso}, and all seem to agree that there is a clear separation of scales, i.e. $\tau_{\text{cloud}}\gg \tau_s$.  For example, in \cite{porto_scales} the lifetime of the cloud has been argued to be as large as:\footnote{\label{footunc} The $\sim$ in \eqref{longlifetime} denotes proportionality constants of order one. We remark that constants of order 1.25 are capable of increasing the average number of squeezed gravitons by an order of magnitude, hence the uncertainty in the upper bound \eqref{N_gr_RESULT_final} and, hence, the estimate \eqref{NgrGR} below.}  
\begin{align}\label{longlifetime}
T\sim 10^7 \tau_s\,,
\end{align}
where $\tau_s$ is the superradiance time scale $\tau_s$, which is given by~\cite{BritoCardoso}:
\begin{equation}
\label{timescale_superradiance}
\tau_s=\frac{1}{\omega_I (2p)}=24\ \mu_b^{-1}\left(\frac{\alpha}{G\mathcal{M}}\right)^{-1}\, a^{ -8}_\mu \,. \end{equation}
In \cite{Dorlis:2025zzz,Dorlis:2025amf}
we consider highly rotating BHs, $\alpha/(G \mathcal{M}) \sim \mathcal{O}(1)$ and $a_\mu=0.1$, consistent with the non-relativistic approximation \cite{BritoCardoso,Detweiller}. 
From \eqref{longlifetime}, \eqref{timescale_superradiance}, we obtain $\mu_b T\sim10^{16}$
and then \eqref{N_gr_RESULT}, implies the following bound interval,
\begin{equation}
   \mathcal{O}(0.1)\leq  \label{N_gr_RESULT_final}
  \sum_{I,J}\vert\mathcal{G}^{(GR)}_{IJ}\vert^2 \lesssim {\mathcal O}(60 - 75) \   ,
\end{equation} 
where the numerical uncertainties in the upper bound are due to the  uncertainties in 
the numerical coefficient (denoted by $\sim$) in \eqref{longlifetime}, as per footnote \ref{footunc}. Thus, it is obvious that the desired exponential enhancement starts to get important in this apparatus, with the large number of ALPs and the large axionic cloud lifetime (interaction time) be the enhancement parameters. Thus, this is the reason that we argue that such an apparatus seems promising on discovering such quantum gravity effects, which means that further enhancement of astrophysical (like a distribution of a large number of black holes, as is the case of the  ``density cusps'' around the Galactic center~\cite{Hailey:2018ocf}) and/or cosmological origin (like squeezing due to inflation~\cite{Kanno:2018cuk,Kanno:2025fpz}) can be realistically considered. We note at this point that, depending on the detailed parameters of the BH and axion masses, one can obtain a range $\sum_{I,J}\vert\mathcal{G}^{(GR)}_{IJ}\vert^2 = {\mathcal O}(10-75)$,
with the upper limit 
\eqref{N_gr_RESULT_final} being pretty robust in the non-relativistic case we consider here.
Hence, on account of \eqref{nu_IJ_expansion} and \eqref{Nmax}, 
the total number of gravitons in the squeezed vacuum state $N_{gr}$ in this case is given by an infinite sum of positive integer powers of quantities of order given typically by \eqref{N_gr_RESULT_final}, which leads to an exponential enhancement 
of $N_{gr}$, 
resembling the behavior \eqref{singlemode} of the single-mode case \cite{Scully_Zubairy_1997}. From \eqref{N_gr_RESULT_final}, therefore, we obtain an estimate of the upper bound of the squeezing parameter 
in our multi-mode squeezed-graviton case
$r_{\rm multi-mode}^2 \equiv     \sum_{I,J}\vert\mathcal{G}^{(GR)}_{IJ}\vert^2 \lesssim {\mathcal O}(60-75)$,
which would correspond to an upper bound on the average number of squeezed gravitons 
\begin{align}\label{NgrGR}
\langle N_{gr}\rangle \lesssim \mathcal O(10^{6}-10^7)\,.
\end{align}
It is important to stress here that the 
observation prospects in the squeezed graviton case at hand are significantly enhanced, as compared to searches that make use of single graviton states~\cite{singlegrav,Carney:2023nzz},
as a consequence of the existence of a macroscopic number of axions in the cloud, due to its condensate nature. 

 In the relativistic case, where curvature effects cannot be avoided, the almost anti-collinear emission of graviton polarisation states breaks down, which might lead to additional enhancement of the squeezing. Par contrast, the CS anomaly-induced squeezing \eqref{N_cs_RESULT} is still highly suppressed, compared to that induced by GR, implying that also in this regime, higher curvature interactions are subleading at low energy scales \cite{Donoghue:1994dn}, being proportional to higher powers of the ration of the axion mass, $\mu_b$, over the Planck mass, while the number of axions cannot compensate such a suppression, due to its linear dependence on the ALPs (coherent) field.

Before closing this session we mention that the non-observation of squeezed single-mode graviton states by LIGO/Virgo interferometers~\cite{LIGOScientific:2016aoc,McCuller:2021mbn} implies~\cite{Hertzberg:2021rbl} an upper bound on the pertinent squeezing parameter 
\be\label{ligosq}
r < 41\,.
\ee
The authors of \cite{Hertzberg:2021rbl} obtained the above result on assuming that only a single-graviton mode, corresponding to a 3-momentum $\vec k^0$, is significantly squeezed in the GW produced from the merger of the two BHs, in the case relevant to the LIGO-Virgo 2015 event~\cite{LIGOScientific:2016aoc}. In Appendix \ref{sec:app} we sketch, for completeness, the proof of this result, so as to clarify the underlying assumptions, which will hopefully help the reader compare (and contrast) that situation with our case.

If the analysis of \cite{Hertzberg:2021rbl}  applied intact to our case, then from \eqref{N_gr_RESULT} one could constrain the axionic-cloud life time, and thus falsify models with too long lifetimes, such as the one in \cite{porto_scales}, leading to the upper bound in \eqref{N_gr_RESULT_final}. However, in our case, the dominant squeezed states are {\it multimode} (polarization and directional) entangled graviton states, making the situation very different from the above. It is not clear whether there are single-mode squeezed graviton states produced from the axion cloud in our scenario, which would allow for the analysis of \cite{Hertzberg:2021rbl} to be applied, thus leading to the bound \eqref{ligosq}. 
We hope to come back to this important issue in a future publication, but in the following subsection, we introduce the basic idea on how to treat the situation in our multi-mode squeezing case. \par 
\section{Detection of quantum features in gravitational waves}\label{sec:qggw}
Observationally, one does not have access to all of the directions of emission from the ALPs cloud. This means, that the relevant phenomenology has to be obtained by tracing out the modes (directions) which one does not have  access to. Since the multi-mode squeezed graviton state has inherent directional entanglement, the tracing out process is a non-trivial task. As we shall demonstrate in this subsection, the tracing out can produce thermal states, and thus thermal noise in an interferometer or even more complex mixed states.

In order to sketch the basic idea, let us consider the pedagogical example of two-mode squeezing, for which the tracing out process leads to exact thermal states.  A two-mode squeezing state is defined by the following operator, 
\begin{align}\label{examplesq2}
S_2(r) = \exp\left[ \xi \left( \hat{a}_1^\dagger \hat{a}_2^\dagger - \hat{a}_1 \hat{a}_2 \right) \right], \quad \xi > 0\,,
\end{align}
where $\xi $ is the squeezing parameter (in general, the squeezing parameters are complex, but in this simplified, but quite relevant to our purposes,  example we take the squeezing parameter to be positive). Then the covariance matrix of the two-mode squeezed state is:
\begin{align}\label{covmatrexample2}
\sigma =
\frac{1}{2}
\begin{pmatrix}
\cosh(2\xi) & 0 & \sinh(2\xi) & 0 \\
0 & \cosh(2\xi) & 0 & -\sinh(2\xi) \\
\sinh(2\xi) & 0 & \cosh(2\xi) & 0 \\
0 & -\sinh(2\xi) & 0 & \cosh(2\xi)
\end{pmatrix}.
\end{align}
Suppose now that the observeer has access only to mode 1, {\it i.e.} one traces out mode 2. The reduced covariance matrix is the upper-left $2 \times 2$ block of the matrix \eqref{covmatrexample2}:
\begin{align}\label{subsystem1}
\sigma_1 =
\frac{1}{2}
\begin{pmatrix}
\cosh(2\xi) & 0 \\
0 & \cosh(2\xi)
\end{pmatrix}
= \frac{\cosh(2\xi)}{2} I_2.
\end{align}
The above covariance matrix \eqref{subsystem1}, pertaining to subsystem 1, is exactly the same as that of a thermal state, with a mean number of gravitons given by $\sinh^2\xi$ ({\it cf.} \eqref{singlemode}). Thus, the relevant phenomenology reduces to that of {\it thermal} states. and is fully determined by the two-mode squeezing parameter of the total (pure) state, before the tracing out process,  and the total number of quanta in it. This means that the entanglement between the modes appears as a {\it thermal noise} when looking at one subsystem. 

We note, for completeness, that the von Neumann entropy of the single-mode 1, which is a Gaussian state \cite{RevModPhys.84.621,Serafini:2017rrn,
adesso2014continuous} with symplectic eigenvalue \( \nu \ge \tfrac{1}{2} \), is:
\begin{align}\label{vNentr}
S(\nu) \equiv \left( \nu + \frac{1}{2} \right) \log \left( \nu + \frac{1}{2} \right)
- \left( \nu - \frac{1}{2} \right) \log \left( \nu - \frac{1}{2} \right).
\end{align}
In our case ({\it cf.} \eqref{subsystem1}):
\begin{align}\label{entrexample2}
\nu = \frac{1}{2} \cosh(2\xi)\,.
\end{align}
Thus, this entropy is zero when $\xi = 0$, and increases with the squeezing parameter $\xi$, reflecting increased entanglement between modes.

In the multi-mode squeezed state case, of interest to us in this talk, the relevant methodology remains the same, albeit with more complex calculations, see \cite{Dorlis:2025amf} and references therein. Reduction to the corresponding (mixed, not pure) single-mode graviton squeezed state, provides the means to calculate the relevant induced noise in the interferometer, and, thus, place constraints on the superradiant ALPs physics. The latter, can be obtained through the upper limits of the relevant squeezed parameter inferred from the currently available LIGO-Virgo-KAGRA data~\cite{LIGOScientific:2016aoc,McCuller:2021mbn}, due to the {\it non-observation} of single-mode squeezed states, according to the analysis of~\cite{Hertzberg:2021rbl}, reviewed in Appendix \ref{sec:app}.\par 
The measurement of the quantum features of GWs that we have just discussed constitutes a challenging step, 
because the latter
appear as noise in the interferometers and not as pure signals~\cite{Parikh:2020kfh}. Independent of quantum effects, the detection of stochasticity in gravitational wave background (SGWB) is important~\cite{Christensen:2018iqi} because it produces a continuous, random and persistent hum due to astrophysical and cosmological sources. The former emanate from incoherent superposition of signals from populations of unresolved sources such as supermassive black holes and binary neutron star mergers. Signals from cosmological sources originate from the very early universe such as those  produced during hypothetical processes like cosmic inflation, phase transitions or cosmic strings. Detecting the SGWB~\cite{Renzini:2022alw} is challenging, because, in a single detector, the signal looks like inherent instrumental noise. The key strategy is to use multiple spatially separated detectors and look for correlated signals that cannot be explained by independent detector noise~\cite{Parikh:2023zat}. The latter is always uncorrelated between detectors. Real gravitational wave signals, traveling across both detectors, will produce a correlated noise in the data streams of the detector pair. In this sense, such noise correlations might be a step forward on circumventing the oppositions of~\cite{Carney:2023nzz,Carney:2024dsj}, and the squeezed graviton states potentially be identified in their detectable regime, e.g. when the squeezing parameter is large enough, since even if each noise can be mimicked on its own by classical GW distributions, their mutual correlation might be not.\par 
Thus, the relevant approach is the one dealing with quantum noise~\cite{Pang:2018eec,Abrahao:2023lle} in gravitational waves. The gravitational states that we have been discussing  are non-classical states. They provide probes for the fundamental nature of spacetime. The nonclassical squeezed states of gravitons are similar to those used in quantum optics \cite{WallsMilburn2008}, where such states are generated in laboratories routinely. Gravitational waves are measured using Michelson interferometers. Modern gravitational wave detectors do not use simple Michelson arms; they have Fabry-Perot (F-P) cavities integrated into each arm. The F-P cavities are formed by the arm mirrors and are designed to increase the effective path length of the laser light. Light is allowed to bounce back and forth thousands of times before exiting the cavity. This multiple pass enhances the interaction time between the light and GW, which proportionally increases the phase shift induced by the gravitational wave strain. In a simple LIGO-like configuration  it has a North arm and East arm of length $L$; a GW stretches one arm and compresses the other and produces opposite phase shifts in the two arm cavities. If we imagine a plus-polarised gravitational wave propagating out of the page, it will stretch the East arm and squeeze the North arm. The Michelson interferometer works by comparing the time it takes for a beam of light to travel down one arm and return versus the time for the light to travel down the other arm and return. The resulting interference pattern at the photodetector depends entirely on the difference between the path lengths. This is the picture of the interaction when the GW is treated classically. A quantum treatment of the interaction between the laser light and gravitational wave has been recently given in~\cite{Pang:2018eec}. Such a coupling, reduces to a coupling of one mode of incident GW with the F-P cavity and the response of the latter is GW state dependent~\cite{Abrahao:2023lle}. Thus, in our case the reduction to the single mode (observed) GW mode and the coupling with a single F-P cavity is the first step forward the detection of the relevant quantum aspects of GWs emitted by ALPs clouds or by any multi-mode squeezed state originated by astrophysical sources, like those presented in~\cite{Kanno:2025how}. Then, one can couple the single mode GW, with multiple F-P cavities and study the correlated noise between them. We hope that such an analysis will be the topic of a forthcoming publication.

\section{The Running-Vacuum-Model Chern-Simons Cosmology and Quantum Gravity}\label{sec:4}

Another important, and related to the above, framework where quantum gravitons play an essential r\^ole, which leads to {\it indirect} detection of QG, is that of the RVM-CS-anomalous inflationary framework, proposed in \cite{Basilakos:2019acj,Mavromatos:2020kzj,Mavromatos:2021urx,Dorlis:2024yqw,Dorlis:2024uei}. According to this scenario, the dynamics of the early Universe are described by a CS gravity action of the form \eqref{CSgravaction}, \eqref{Rcs}, stemming from string theory, in which case the $b(x)$ field is the {\it massless} string-model independent KR axion~\cite{Svrcek:2006yi}, and the coupling $A$ of the CS interaction is given by \eqref{Adef}. That is, the matter action \eqref{matter} is characterized in this case by $\mu_b=0$, in contrast to our astrophysical BH case discussed previously. In the absence of gCS anomaly terms, {\it i.e.} when $\mathcal S_{CS}=0$, the action \eqref{CSgravaction}, \eqref{matter}, with $\mu_b=0$, without further self-interactions of the $b$ field, describes a cosmology with a stiff KR axion era. In the model of \cite{Basilakos:2019acj,Mavromatos:2020kzj,Mavromatos:2021urx} such an era precedes an inflationary era. Due to its stringy origin and its links to RVM cosmology, this cosmological model is termed as Stringy RVM or StRVM for short.

This universe may be characterized~\cite{Mavromatos:2020kzj} by the presence of {\it chiral}
primordial GWs at the end of the stiff era. There is a plethora of sources of such GWs in the primordial universe~\cite{Mavromatos:2020kzj}, including non spherical collapse of appropriate (biased) domain walls, or merging of primordial BHs. The GWs 
can condense yielding a running-vacuum-model (RVM)~\cite{Lima:2013dmf,SolaPeracaula:2022hpd} type inflation, which does not require fundamental inflaton fields. The inflation in such models is due to the non-linear dependence on the Hubble parameter of the condensate of the gCS terms~\cite{Basilakos:2019acj,Dorlis:2024yqw}, \eqref{Rcs}. The latter can be estimated in the weak-graviton approximation upon expanding about a Friedman-Lemaitre-Robertson-Walker (FLRW) spacetime background as in \eqref{metrpert}. On restricting the effective gravitational field theory expansion to quadratic GW perturbations, which suffices in a weak-graviton approximation, one can compute the leading QG contributions to the gCS condensate. The condensate has real and imaginary parts~\cite{Dorlis:2024uei}, the latter indicating the instability of the StRVM inflationary ground state induced by the former. 

Assuming isotropic and homogeneous dependence of all fields, the real parts of the gCS condensate are given by~\cite{Dorlis:2024yqw}:
\begin{align}\label{gCScond}
    \langle A\, \mathcal R_{\rm CS} \rangle^I_{\mathcal N_I} = -\mathcal{N}_I\frac{ A^2\,  \kappa ^4 \mu ^4}{\pi ^2}\dot{b}_I H_{I}^3\,,
\end{align}
where $H_I$ is the approximately constant Hubble parameter, during the induced inflationary era, $\mu=M_s$ denotes the UltraViolet (UV) cutoff of the effective gravitational field theory embedded in strings, and $\mathcal N_I$ is the number of GW sources at the onset of the RVM inflation. The quantity 
\begin{equation}
    \dot{b}_I\sim 10^{-1} H_I M_{\rm Pl}\,,
    \label{bdotInflation}
\end{equation}
is the (approximately) constant rate of change of the KR axion during the inflationary era. This rate can be justified using a dynamical-system approach to inflation in this model~\cite{Dorlis:2024yqw}. 

\begin{figure}[ht!]
    \centering
\includegraphics[width=0.9\textwidth]{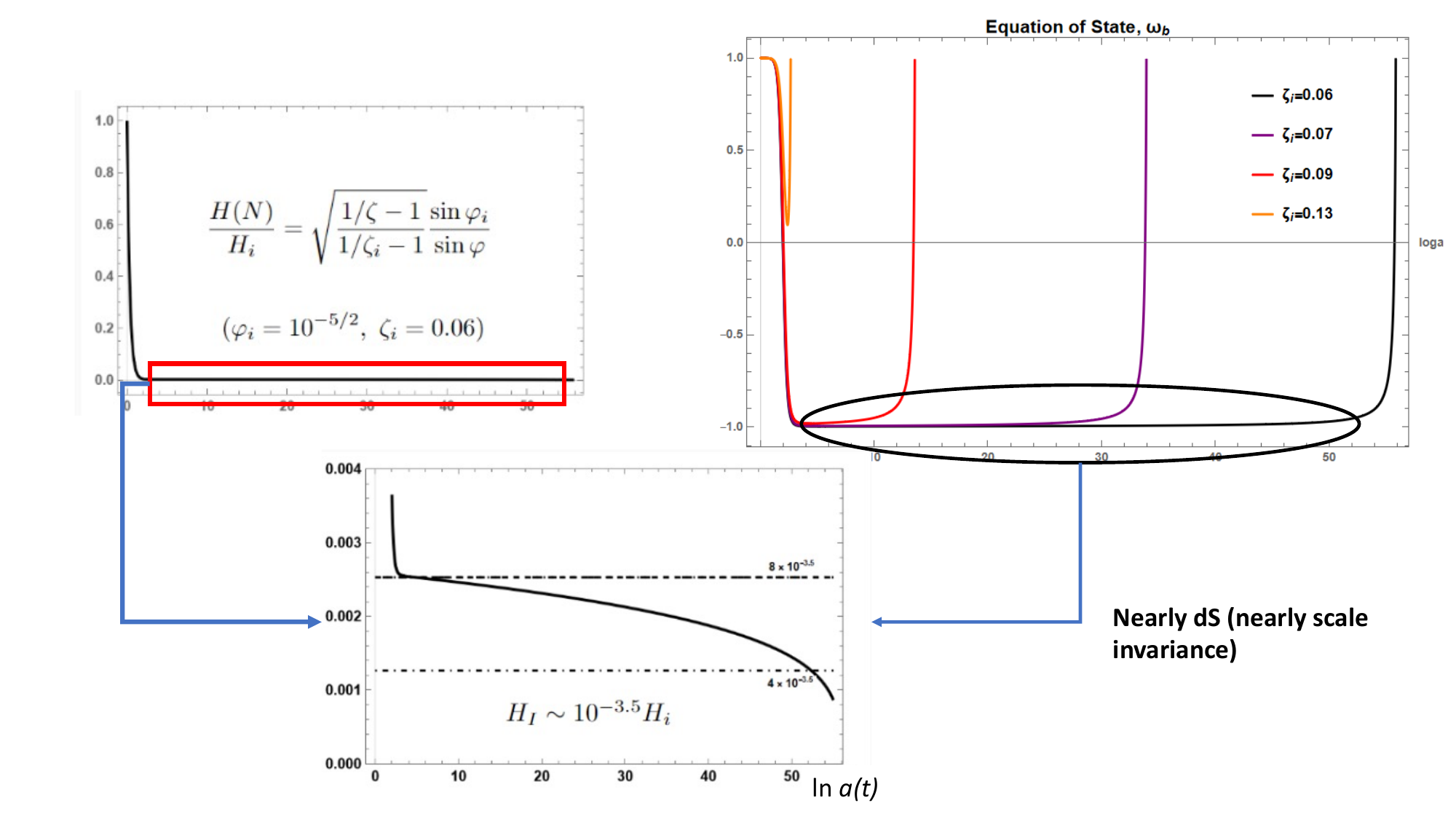}
    \caption{The evolution of the Hubble parameter during the short period from the end of the axion-dominated stiff era to the inflationary phase, in the stringy RVM cosmology of \cite{Basilakos:2019acj,Mavromatos:2020kzj,Dorlis:2024yqw}.
    The Hubble rate drops by almost four orders of magnitude during that transition. Figure taken from \cite{Mavromatos:2025mmo}.}
    \label{fig:HEvol}
\end{figure}

The real part of the condensate \eqref{gCScond} leads to an approximately linear potential for the axion $b(x)$ 
\begin{align}\label{potb}
   V(b) = b(x) \, A\langle \, \mathcal R_{\rm CS}\rangle^{I}_{\mathcal N_I} \equiv b(x) \, \mathcal C \,,
\end{align}
with $\mathcal C <0$ is approximately constant given by \eqref{gCScond}, provided $H_I$ is approximately constant. The dynamical system analysis of \cite{Dorlis:2024yqw,Dorlis:2024uei} leads to an inflationary equation of state for a duration that is determined by the imaginary parts of the condensate~\cite{Dorlis:2024uei}
\begin{align}\label{imRCS}
 {\rm Im}\langle \widehat{R_{\mu\nu\rho\sigma}\, \widetilde R^{\nu\mu\rho\sigma}} \rangle = 
 \frac{16  A \text{$\dot{b}$} \mu ^7}{7 M^{4}_{\rm Pl} \ \pi ^2} \left[1+\left(\frac{H_I}{\mu}\right)^2\left(\frac{21}{10}-6\left(\frac{A\mu\text{$\dot{b}$}}{M_{\rm Pl}^2}\right)^2\right)\right] \ . 
\end{align}
The induced lifetime is proportional to such imaginary parts, as explained in \cite{Dorlis:2024uei}. Requiring the inflationary life time $\tau_{\rm infl}$ to be consistent with the cosmological data~\cite{Planck}, that is 
of order $\tau_{\rm infl} = \mathcal O(50-60)\, H_I^{-1}$, with
\begin{align}\label{HIMP}
H_I \lesssim 10^{-5}\, M_{\rm Pl}\,, 
\end{align}
yields a constraint on the string scale \begin{equation}\label{MsMP}
\frac{M_s}{M_{\rm Pl}}\lesssim
0.215\,,
\end{equation}
which is consistent with the results of the linear-axion-potential dynamical-system analysis of \cite{Dorlis:2024yqw}, given that life times of such order are included in the initial conditions for inflation in the dynamical system analysis, as becomes clear from figure~\ref{fig:HEvol}, which depicts the evolution of the Hubble and the equation of state in the model of \cite{Basilakos:2019acj,Mavromatos:2020kzj,Dorlis:2024yqw} during the onset of the RVM inflation from a preceding stiff axion-dominated era~\cite{Mavromatos:2025mmo}. 
The analysis of \cite{Dorlis:2024yqw} demonstrates that, during the onset of inflation the value of the Hubble rate drops significantly (by almost four orders of magnitude), at the expense of a sudden increase of the primordial GW sources. Indeed, in general we have:
\begin{equation}
 \frac{H(N)}{H_i}= \sqrt{\frac{1/\zeta-1}{1/\zeta_i-1}}\frac{\sin\phi_i}{\sin\phi}\,,
\end{equation}
where the subscript $i$ denotes initial values, at the end of the stiff axion-dominated era, and the onset of RVM inflation. The parameters $\zeta_i, \phi_i$ are defined in the dynamical system approach
to inflation in \cite{Dorlis:2024yqw}.
For the specific initial conditions leading to inflation:
$\phi_i=10^{-5/2},\, \,\zeta_i=0.06$,
the evolution of the Hubble parameter $(t)$ is depicted in figure~\ref{fig:HEvol}. Such initial values, yield  
\begin{align}\label{HIi}
H_I\sim 10^{-3.5}H_i\,.   
\end{align}
Agreement with cosmological data~\cite{Planck} requires
\begin{align}\label{HI}
H_I \lesssim 10^{-5}\, M_{\rm Pl}\,.
\end{align}
From the result \eqref{HIi}, we have that, in order to guarantee a smooth passage from the stiff era to the anomaly-dominated condensate inflation,
the value of the condensate \eqref{gCScond} at the onset of inflation should be the same with that at the end of the stiff era, characterized by $\mathcal N_S$ sources of GWs. During the end of the stiff era, the (real part of the) gCS condensate can be estimated as~\cite{Dorlis:2024yqw}:
\begin{equation}\label{condstiff}
\langle A\, \mathcal R_{\rm CS}\rangle^{\rm stiff}_
      {\mathcal N_S}= -\mathcal N_S \, \frac{30\sqrt{6} A^2\, \kappa ^3 \mu ^4}{\pi ^2} \, H_i^4\,.
\end{equation}
Equating \eqref{condstiff} with \eqref{gCScond}, and taking into account \eqref{HIi}, we finally obtain:
\begin{align}\label{sourcesNINS}
    \frac{\mathcal{N}_I}{\mathcal{N}_S}\sim 7\cdot 10^{16}\,.
\end{align}
The result is consistent with keeping the UV cutoff at sub-Planckian regimes, given that $\mu = M_s < M_{\rm Pl}$, as a result of \eqref{MsMP}. Using the expressions~\cite{Dorlis:2024yqw}
$\frac{b}{M_{\rm Pl}}=\frac{\zeta - 1}{\zeta}$ and $\frac{\dot{b}}{H M_{\rm Pl}}= \sqrt{6} \cos\phi$, 
which are derived in the dynamical-system approach to inflation~\cite{Dorlis:2024yqw,Dorlis:2024uei}, 
one can estimate that, for the above-described boundary conditions leading to inflation, one arrives at \eqref{bdotInflation}.mentioned previously. 

The above define the conditions required for a condensate inflation to occur. After that drop, the Hubble parameter remains approximately constant during the finite lifetime of the metastable, approximately de Sitter (dS), RVM inflationary period, determined, as discussed above, by the imaginary parts of the gCS condensate \eqref{imRCS}.
Although triggered by the approximately linear potential for the massless KR axion, nonetheless it is an RVM-type non linear inflation, not requiring massive inflaton fields. The linear axion potential only assists the inflation.

However, the presence of such a potential, which is purely quantum in origin, due to the anomalous gCS condensate, amplifies effects of periodic modulations of the axion potential at early epochs, due to non-perturbative instanton effects in the gauge sector of the model~\cite{Dorlis:2025gvb}, on the profile of GW in the early radiation era after inflation~\cite{Mavromatos:2022yql}.
Thus, although in this case, a direct detection of quantum gravitons is not applicable, nonetheless QG plays an important role, as we discussed above, on the formation of the gravitationally anomalous CS condensate, which creates an RVM type metastable inflation, with all the non trivial phenomenological consequences the latter implies~\cite{SolaPeracaula:2022hpd}. These 
include prolonged reheating~\cite{Lima:2013dmf}, with perhaps early dominated matter eras, 
leading to enhanced production of primordial black holes in the model~\cite{Papanikolaou:2024rlq,Tzerefos:2024rgb}, with implications on the profile of GWs in early radiation eras, 
as well as alleviations~\cite{Gomez-Valent:2023hov} of the observed 
cosmological tensions in the current era~\cite{CosmoVerseNetwork:2025alb}.

On the other hand, if there exist populations of rotating primordial BHs during the inflationary era, 
it is possible that their mergers produce squeezed GW states, according to similar mechanisms discussed in section \ref{sec:2}. In such a case, squeezed graviton states can be produced during inflation, which can leave, in principle, depending on the properties of the primordial BHs - observable traces in the profiles of GWs after inflation. Such a production of squeezed graviton states can also enhance the production of squeezed states during inflation according to standard arguments~\cite{Grishchuk:1989ss,Grishchuk:1990bj,Albrecht:1992kf,Mukhanov:2007zz,Kanno:2018cuk,Kanno:2019gqw,Kanno:2021vwu,Kanno:2022ykw}. Since, due to environmental entanglement, such inflationary-era-produced squeezed states will eventually decohere, these states might be detected indirectly, {\it e.g.} 
through a method proposed in \cite{Kanno:2021gpt}, based on the observation of the decoherence time of the entangled state, as a result, for instance, of coupling of gravitons to cosmic photons, which is one source of environmentally-induced decoherence. Similar methods might be applicable to our astrophysical-BH-induced squeezing, discussed in section~\ref{sec:2}.

\section{Conclusions and Outlook}\label{sec:5}

We have discussed aspects of perturbative QG, viewing the latter as an effective field theory, with weak graviton perturbations around non-trivial spacetime backgrounds, either rotating black holes or cosmological Friedman-Lemaitre-Robertson-Walker Universes in the presence of primordial chiral gravitational waves.
In both frameworks quantum gravity effects were essential in ensuring a non-trivial r\^ole of gravitational anomalies coupled to ALPs. The underlying effective field-theoretical framework is semi-classical Chern-Simons gravity. 

In the presence of a rotating black hole, the ALPs, if massive, can form condensates (``clouds'') in the exterior of the black hole, in which interactions of ALPs produce, under some circumstances, a significant number of quantum-entangled pairs of squeezed gravitons, which might be detectable in future interferometric devices. At present, the non observation of such squeezed states by currently operating interferometers (LIGO-Virgo-KAGRA) imposes upper bound constraints on the relevant squeezing parameters, and thus the lifetime of the axionic clouds. 

In the early Universe cosmological frameworks, on the other hand, chiral quantum GW perturbations can condense in the primordial Universe, leading in turn to a condensate of the gCS anomaly terms.
This in turn leads to inflation of RVM type, with highly non-trivial phenomenological consequences, such as prolonged reheating and deviating behaviour from $\Lambda$CDM paradigm at modern epochs, including the potential alleviation of the cosmological tensions, as well as modified profiles of GW in the early radiation era, detectable by future interferometers, such as LISA~\cite{Berti:2005ys,Caprini:2015zlo,Brito_2017_scales,Caprini:2019egz}. 
The latter framework and its consequences would therefore constitute indirect evidence for quantization of gravity, in contrast to the former framework, which, if realized in nature, would directly falsify or verify the quantum nature of the gravitational interaction. 

Nonetheless, production of a sizable amount of squeezed states of gravitons according to the mechanisms proposed in section \ref{sec:2} can also take place during inflation, if, for instance, there are appropriate populations of rotating black holes (surrounded by axionic clouds) created from the inflationary vacuum through quantum fluctuations. Such squeezed state can enhance the squeezed states of gravitons produced during inflation through standard arguments.

Finally, we mention that one of the future research directions would be the exploration of how the quantum state of the gravitational field—beyond its classical waveform—may become experimentally accessible through precision interferometry. In this respect, we mention that the detector itself can be formulated as an open quantum system, in which the optical field, test-mass motion, and gravitational strain are treated within a unified Hamiltonian framework. From this one can deduce that nonclassical features of the gravitational field are in principle transferred to the optical output through well-defined input--output channels.

This raises the intriguing possibility that future interferometers, operating deep in the quantum backaction regime and augmented by squeezed-light techniques and long-baseline correlations, could become sensitive not only to the energy carried by gravitational waves but also to their quantum correlations, squeezing, or even entanglement. In this sense, interferometric detectors may ultimately function as macroscopic quantum transducers linking microscopic models of graviton production—such as those arising from axion clouds, black-hole superradiance, or early-Universe dynamics—to directly observable optical signatures. While formidable experimental challenges remain, the framework developed here suggests that precision gravitational-wave astronomy may one day evolve into a genuine arena for testing the quantum nature of spacetime itself.
We hope to come back to a detailed discussion of the above important issues in a future publication.

\section*{Acknowledgments}

NEM would like to thank the organisers of 
the Corfu Summer Institute 2025 Workshops 
{\it on the Standard Model and Beyond}
(Aug 24 - Sep 03, 2025), and 
{\it Tensions in Cosmology} (Sep 02 - Sep 08, 2025), for the invitation to speak and for organising thought stimulating and high quality meetings.
The work of P.D. is supported by a graduate scholarship from the National Technical University of Athens (Greece).
The work of NEM and SS is supported in part by the UK Science and Technology Facilities research Council (STFC) under the research grant No. ST/X000753/1, and by the UK Engineering and Physical Sciences Research Council (EPSRC) under the research grant No. EP/V002821/1. 
The work of S.-N.V. is supported by the Hellenic Foundation for Research and Innovation
(H.F.R.I. (EL.ID.EK.)) under the “5th Call for H.F.R.I. Scholarships to PhD Candidates” (Scholarship Number:
20572).
NEM also acknowledges participation in the COST Association Actions CA21136 “Addressing observational
tensions in cosmology with systematics and fundamental physics (CosmoVerse)” and CA23130 ``Bridging high and
low energies in search of quantum gravity (BridgeQG)”.

\appendix

\section{Non-Observation of single-mode squeezed gravitons by current interferometers }\label{sec:app}

 In a weak field expansion about Minkowski spacetime of the gravitational field, as appropriate for the measurements in the interferometers, the corresponding Fourier transform of the GW perturbation with respect to the 3-momentum $\vec k$ reads:
\begin{align}\label{hft}
\widetilde h_{\vec k, p}(t) = \int d^3 x \, h_p(\vec x, t)  \, \exp\Big(-i \vec k \cdot \vec x\Big)\,,
\end{align}
where $h_p $ denotes the two GW polarizations 
$h_p (\vec x, t) \equiv \Big( h_\times, \, h_{+} \Big)$,
and we have absorbed, for convenience, the 
gravitational coupling $\kappa$  in the definition of the tensor perturbation, so that $h_p$ is dimensionless.
In canonical quantization, we may enclose the system in a finite volume, which discretizes the momentum integrals. 
Following the normalization of \cite{Hertzberg:2021rbl} the energy of the GW can be expressed as:
\begin{align}\label{discrener}
E &= \sum_{p=\times,+} \sum_{\vec k} E_{\vec k,p}\, \frac{1}{4\kappa^2}\,
\Big( \vert \dot{\widetilde{h}}_{\vec k,p}\vert^2 + k^2\, \vert {\widetilde{h}}_{\vec k,p}\vert^2 \Big)\,.
\end{align}
Comparing the expression for the individual-mode energies $E_{\vec k,p}$ in \eqref{discrener} with the energy of the simple harmonic oscillator (SHO), one observes that the modes ${\widetilde h}_{\vec k,p}(t)$
play the r\^ole of the spatial positions $x(t)$ of the SHO. 

Viewing GW as a collection of harmonic oscillators, in which the position $x$ is replaced by the appropriate Fourier modes of the graviton tensor perturbations of polarizartion $p$, 
momentum $\vec k$, and frequency $\omega = \vert \vec k \vert \equiv k$, the 
squeezed single-mode GW function can be constructed in full analogy with the SHO case:
\begin{align}\label{GWsq}
\psi_s(h,t) &\propto \prod_{a=1,2}\, \prod_{\vec k} \, \prod_{p=\times,+} \, \exp \Big(i\, \epsilon_{a \vec k, p}(t) + \frac{i}{2\hbar} \, \pi_{c, a \vec k, p}\, \widetilde{h}_{a \vec k, p} - \frac{k\, S_{a \vec k, p}(t)}{8\kappa^2 \hbar V} \, (\widetilde{h}_{a \vec k, p} - \widetilde{h}_{c, a \vec k, p})^2 \Big)\,, \nonumber \\
\epsilon_{a \vec k, p}(t) &=  -\frac{k}{4}\, \int_0^t dt^\prime \, S_{a \vec k , p}(t^\prime) - \frac{1}{4\hbar} \widetilde{h}_{c, a \vec k, p}(t) \, \pi_{c, a \vec k, p}(t)\,, \nonumber \\
S_{a \vec k, p}(t) &= \tanh \Big(\tanh^{-1}(\beta_{a \vec k, p}) + i k\, t\Big), \nonumber \\
\end{align}
with the boundary condition 
for the squeezing function at $t=0$: 
$S_{a \vec k, p} (0) =\beta_{a \vec k,p}$, where 
$\beta_{a \vec k,p}$ are the squeezing parameters. The  index $a=1,2$ denotes the real ($a=1$) and imaginary ($a=2$) parts of the complex 
quantity $\widetilde{h}_{\vec k, p}$, in the compact notation of \cite{Hertzberg:2021rbl}. The 
$\widetilde{h}_{c, a \vec k, p}(t)$ satisfy the classical equations of motion (which, for each GW mode, is quite analogous to the SHO equation for the spatial positions $x(t)$): 
\be\label{GWeqs}
\ddot{\widetilde h}_{c, a \vec k, p} + k^2 \, \widetilde{h}_{c, a \vec k, p} =0\,,
\ee
and $\pi_{c, a \vec k, p}(t)$ are the canonical momenta conjugate to $\widetilde{h}_{c, a \vec k, p}(t)$. As in the case of the SHO, upon setting $S_{a \vec k, p}(t) = 1$, the squeezed state wavefunction 
\eqref{GWsq} corresponds  to a coherent graviton wavefunction.
For the reader it is important to realize the physical mechanisms behind the formation of squeezed single-mode states. At a given point in time, here taken to be a common one $t=0$, the frequency of each GW mode is rescaled by the corresponding squeezing parameter:
\be\label{freqshifts}
k \to \beta_{a\vec k, p} \, k\,.
\ee
What causes such a squeezing microscopically is not discussed in the phenomenological model of \cite{Hertzberg:2021rbl}. In our rotating-BH-axion-cloud context, such a single-mode squeezing could be the result of collective interactions of gravitons with ALPs, but it is not clear if such processes actually take place in our situation. This should be investigated further elsewhere.

The authors of \cite{Hertzberg:2021rbl} constructed single-mode squeezed states in the simplified example 
of two identical squeezing functions 
\be\label{idsqfunc}
S_{\vec k, p}(t) = S_{1 \vec k p}(t) = S_{2 \vec k, p}(t)\,.
\ee
For the interferometric measurements, the important quantities in the analysis of \cite{Hertzberg:2021rbl} are the correlators of the deviations from the mean, $\delta h_p \equiv h_p - \langle h_p\rangle$, where the $\langle \dots \rangle$ is taken with respect to the single-model GW squeezed state \eqref{GWsq}. The experimentally relevant quantity reads:
\begin{align}\label{xidef}
\xi_p(\vec x, \vec y, t, t^\prime) &\equiv \langle \delta h_p (\vec x, t) \, \delta h_p (\vec y, t)\rangle 
= \langle h_p(\vec x, t) \, h_p(\vec y, t^\prime) - \langle h_p (\vec x, t)\rangle \, \langle h_p (\vec y,t ^\prime) \rangle \nonumber \\ &= \int \frac{d^3k}{(2\pi)^3} \, \frac{\kappa \hbar}{k} \, \Big(\beta_{\vec k, p}^{-1} \, \cos (k\, t) \, \cos (k \, t^\prime) + \beta_{\vec k, p} \, \sin (k\, t) \, \sin (k \, t^\prime) + i \sin (k\, (t^\prime - t))\Big)\,.
\end{align}
On the assumption that only a single-mode with momentum $\vec k^0$ is significantly squeezed, one can write~\cite{Hertzberg:2021rbl}
\be\label{singlemodeb}
\beta_{\vec k, p} = 1 + \frac{e^{2\, r_p} \, k^{0\,3}}{2}\, (2\pi)^3 \Big(\delta^{(3)}(\vec k - \vec k^\prime) + \delta^{(3)}(\vec k + \vec k^\prime)\Big)\,,
\ee
where $r_p$ is a dimensionless parameter, denoting the squeezing strength, and the above expression guarantees the reality condition for the squeezing parameter 
$\beta_{\vec k, p} = \beta^\star_{-\vec k, p}$. 
The authors of \cite{Hertzberg:2021rbl} argued in favour of a physical extension of the above monochromatic approximation by replacing the squeezing parameter \eqref{singlemodeb} with a smoothened out expression, involving the replacement of the $\delta^{(3)} (\vec k - \vec k^0)$ functions by Gaussian distributions, peaked at $k^0$, with width $\sigma_0^2$:
\begin{align}\label{betasmoothened} 
\beta_{\vec k, p} = 1 + \frac{e^{2r_p}\, k^{0\,2}\, k}{2} \delta(k_x) \delta(k_y) \, \frac{1}{\sqrt{2\pi\,\sigma_0^2}} \Big[ \exp \Big(-\frac{(k_z - k^0)^2}{2\sigma_0^2}\Big) +  exp \Big(-\frac{(k_z + k^0)^2}{2\sigma_0^2}\Big)\Big]\,,
\end{align}
for the concrete example of a GW propagating along the (positive) z direction, with mean wavenumber $\vec k^{0} = k^0 \, \widehat z$, with  $\widehat{z}$ denoting unit vector along the $z$ direction. The factor $k^{0\,3}$ in 
\eqref{singlemodeb} has also been smoothened out in a particular way, for convenience.
Using then \eqref{betasmoothened}, one obtains the regularized $\xi_p(\vec x, \vec y, t, t^\prime)$ correlator \eqref{xidef}:
\begin{align}\label{regxi}
\xi_p^{\rm reg}(\vec x, \vec y, t, t^\prime) &= \frac{k \hbar}{2\pi^2 (\vert \vec x - \vec x^\prime\vert^2 - (t - t^\prime)^2)}
\nonumber \\ &+
\frac{\kappa \hbar}{4}\, \mu_p (z,t) \, e^{2r_p} \, k^{0\,2} \sum_{\mp} e^{-( (z-z^\prime) \mp (t-t^\prime))\, \sigma_0^2/2} \cos(k^0 ((z-z^\prime) \mp (t-t^\prime)))\,,
\end{align}
where the factor $\mu_p(z,t)$
multiplying the second term in \eqref{regxi} is a modulating factor of the form:
\begin{equation}\label{modfunct}
\mu_P(z,t) =
\begin{cases}
1, & \lvert z - t - \phi_c \rvert \ll \lambda_c, \\
0, & \lvert z - t - \phi_c \rvert \gg \lambda_c\,.
\end{cases}
\end{equation}
which has been inserted by the authors of \cite{Hertzberg:2021rbl}, beyond the regularization of the squeezing parameter \eqref{betasmoothened}, in order to ensure that the contributions of the second term of \eqref{regxi} are appreciable only at the centre of the classical wavepacket, $h_c$, which has a size $\lambda_c$ and its centre has a phase $\phi_c$. 

Using the above Gaussian wavepacket 
representation of the monochromatic squeezed graviton state, the authors of \cite{Hertzberg:2021rbl}
have calculated the detector's response to such a squeezing state.
Ignoring the coherent-graviton part, given that the latter resembles a classical state, and therefore its fluctuations will be suppressed, compared to the squeezed state, one can concentrate on the second term in the regularised correlator \eqref{regxi}, namely:
\begin{align}\label{detectorregxi}
\langle \delta h_p (t) \, \delta h_p (t^\prime) \rangle^{\rm reg} = \frac{\kappa \hbar}{2} \, e^{2r_p}\,k^{0\,2} \, \mu_p(z,t) \, e^{-(t-t^\prime)\, \sigma_0^2/2}\, \cos\Big(k^0\, (t-t^\prime)\Big)\,.
\end{align}
On evaluating the amplitude of the fluctuations at each moment of time, {\it i.e.} taking the limit $t \to t^\prime$, and concentrating on the maximum fluctuation, that is setting $\mu_p \to 1$, one readily obtains from \eqref{detectorregxi} the standard deviation of a squeezed graviton state~\cite{Hertzberg:2021rbl}:
\begin{align}\label{standarddevsq}
\sigma_S & \equiv \sqrt{\langle (\delta h_p)^2\rangle} = \sqrt{4\pi} \, e^{r_p} \frac{2\pi\,f^0}{\omega_{\rm Pl}}\, , \nonumber \\
f^0 &= \frac{k^0}{2\pi} \, ,  \quad \omega_{\rm Pl} \equiv \sqrt{\frac{8\pi}{\hbar}}\, \frac{1}{\kappa} \simeq 1.9 \times 10^{-43} \, \rm s^{-1} \,\, (``\rm Planck~freqwuency'')\,.
\end{align}
Taking into account that the standard deviations from LIGO-Virgo~\cite{LIGOScientific:2016aoc,McCuller:2021mbn} at Livingston (L) and Hanford (H) are:
\begin{align}\label{sdev}
\sigma_L \simeq \sigma_L \simeq 1.6 \times 10^{-22}\,, 
\end{align}
and $f^\star \simeq 200 ~\rm Hz$, and interpreting the lack of observations of squeezed GW states at present as implying
$\sigma_s < \sigma_L \simeq \sigma_H $, 
the authors of \cite{Hertzberg:2021rbl}
arrived, on account to \eqref{sdev}, to the following conservative bound on the squeezing strength $r_p$:
\begin{align}\label{boundsqstr}
r_p < 41\,,
\end{align}
which is the announced result \eqref{ligosq}.

\bibliographystyle{apsrev4-2}
\bibliography{squeezing_Corfu2025.bib}

\end{document}